\documentclass{iopart}

\usepackage{ifthen}
\usepackage{ifpdf}
\usepackage{latexsym}
\usepackage{amssymb}
\usepackage{bm}

\ifpdf
\usepackage{graphicx}
\usepackage{epstopdf}
\else
\usepackage{graphicx}
\usepackage{epsfig}
\fi

\def\Xint#1{\mathchoice
   {\XXint\displaystyle\textstyle{#1}}%
   {\XXint\textstyle\scriptstyle{#1}}%
   {\XXint\scriptstyle\scriptscriptstyle{#1}}%
   {\XXint\scriptscriptstyle\scriptscriptstyle{#1}}%
   \!\int}
\def\XXint#1#2#3{{\setbox0=\hbox{$#1{#2#3}{\int}$}
     \vcenter{\hbox{$#2#3$}}\kern-.5\wd0}}

\def\dashint{\Xint-}

\newcommand{\im}{\mbox{Im}}

\newcommand{\eexp}{\mbox{e}^}

\newcommand{\tbox}[1]{\mbox{\tiny #1}}

\newcommand{\be}[1]{\begin{eqnarray}\ifthenelse{#1=-1}{\nonumber}{\ifthenelse{#1=0}{}{\label{e#1}}}}
\newcommand{\ee}{\end{eqnarray}} 

\newcommand{\hide}[1]{}

\newcommand{\sect}[1]{\section{#1}}

\begin{document}

\title[Quantum decay]{Quantum decay into a non-flat continuum}

\author{James~Aisenberg$^{1}$, Itamar~Sela$^{2}$, Tsampikos~Kottos$^{1}$, Doron~Cohen$^{2}$, Alex~Elgart$^{3}$}

\address{
\mbox{$^1$Department of Physics, Wesleyan University, Middletown, CT 06459, USA}\\
\mbox{$^2$Department of Physics, Ben-Gurion University, Beer-Sheva 84105, Israel}\\
\mbox{$^3$Department of Mathematics, Virginia Tech, Blacksburg, VA 24061, USA}}

\begin{abstract}
We study the decay of a prepared state into non-flat continuum.
We find that the survival probability $P(t)$ might exhibit either
stretched-exponential or power-law decay, depending on non-universal
features of the model. Still there is a universal characteristic
time $t_0$ that does not depend on the functional form.
It is only for a flat continuum that we get a robust exponential
decay that is insensitive to the nature of the intra-continuum couplings.
The analysis highlights the co-existence of perturbative and
non-perturbative features in the local density of states,
and the non-linear dependence of $1/t_0$ on the strength of the coupling.
\end{abstract}

%%%%%%%%%%%%%%%%%%%%%%%%%%%%%%%%%%%%%%%%%%%%%%%%%%%%%%%%%%%%%%%%%%%%%%%%%%%%%%%%%%%%%%%%%%%%%%%%%%%%
%%%%%%%%%%%%%%%%%%%%%%%%%%%%%%%%%%%%%%%%%%%%%%%%%%%%%%%%%%%%%%%%%%%%%%%%%%%%%%%%%%%%%%%%%%%%%%%%%%%%
\sect{Introduction}

The time relaxation of a quantum-mechanical prepared state
into a continuum due to some residual interaction 
is of great interest in many fields of physics. 
Applications can be found in areas as diverse as 
nuclear \cite{AZ02}, 
atomic and molecular physics \cite{CDG92} 
to quantum information \cite{NC00}, 
solid-state physics \cite{PAEI94,BH91}
and quantum chaos \cite{PRSB00}. 
The most fundamental measure characterizing
the time relaxation process is the so-called survival probability~$P(t)$,
defined as the probability not to decay before time~$t$.

The study of $P(t)$ goes back to the work
of Weisskopf and Wigner \cite{WW30} regarding 
the decay of a bound state into a flat continuum.
They have found that~$P(t)$  follows an exponential 
decay ${P(t)=\exp(-t/t_0)}$, with a rate $1/t_0$
which is given by the Fermi Golden Rule (FGR), 
and hence proportional to the effective density 
of states (DOS) for ${\omega=0}$ (energy conserving) transitions.

Following Wigner, many studies have adopted
Random Matrix Theory (RMT) modeling \cite{izrailev,fyodo}
for the investigation of~$P(t)$, 
highlighting the importance of the statistical properties 
of the spectrum \cite{GACMP97}.  
Notably in the context of a many-particle system, one should
understand the role of the whole hierarchy of states and associated
couplings, ranging from the single-particle levels to the
exponentially dense spectrum of complicated many-particle
excitations \cite{S01}, e.g., leading to a decay ${P(t) \sim \exp(-{\sqrt t})}$.
Non-uniform couplings also emerge upon quantization 
of chaotic systems where non-universal (semiclassical) features
dictate the band-structure of the perturbation, leading to a highly
non-linear energy spreading \cite{brm}.

%%%%%%%%%%%%%%%%%%%%%%%%%%%%%%%%%%%%%%%%%%%%%%%%%%%%%%%%%%%%%%%%
%%%%%%%%%%%%%%%%%%%%%%%%%%%%%%%%%%%%%%%%%%%%%%%%%%%%%%%%%%%%%%%%
{\bf Motivation. --}
Despite all the mounting interest in physical circumstances 
with complex energy landscape,  
a theoretical investigation of the time relaxation
for prototypical RMT models is still missing, and also
the general (not model specific) perspective are lacking.
A reasonable starting point for an RMT modeling 
is the characterization of the physical system 
of interest by a spectral function $\tilde{C}(\omega)$  
that describes the power spectrum of its fluctuations
(the exact definition is given in the next section).
For an idealized strongly chaotic systems this power 
spectrum looks ``flat", or using an optional terminology 
taken from different context it is called ``white" or ``Ohmic".  
But in more realistic circumstances  $\tilde{C}(\omega)$ 
is not flat (see some examples in \cite{brm,dil}), 
and one wonders what are the consequences. 
Of particular interest are circumstances 
in which for small frequencies 
${\tilde{C}(\omega) \propto \omega^{s{-}1}}$
with $s<1$ (``sub-Ohmic" spectral function) 
or $s>1$ (``super-Ohmic" spectral function).
For such extreme non-flatness the conventional 
Wigner-Weisskopf-FGR picture is not applicable, 
giving zero or infinite rate of decay respectively. 
For this reason the decay into an ${s\ne1}$ continuum 
is the most interesting and challenging case for analysis.

%%%%%%%%%%%%%%%%%%%%%%%%%%%%%%%%%%%%%%%%%%%%%%%%%%%%%%%%%%%%%%%%
%%%%%%%%%%%%%%%%%%%%%%%%%%%%%%%%%%%%%%%%%%%%%%%%%%%%%%%%%%%%%%%%
{\bf Scope. --}
In this paper, we explore a general class of prototype models where the initial
state decays into a non-flat (sub-Ohmic or super-Ohmic) continuum. 
%
% In the language of quantum dissipation
% studies, this means that we are dealing with a non-Ohmic model.
%
We show that the survival probability $P(t)=g(t/t_0)$ is characterized
by a generalized Wigner decay time $t_0$ that depends in a non-linear
way on the strength of the coupling. We also establish that the
scaling function $g$ has distinct universal and
non-universal features. It is only for the flat continuum
of the traditional Wigner model, that we get a robust
exponential decay that is insensitive to the nature of the intra-continuum couplings.
In addition to $P(t)$ we investigate other characteristics of the evolving
wavepacket, namely the variance $\Delta E_{\tbox{sprd}}(t)$
and the $50\%$ probability width $\Delta E_{\tbox{core}}(t)$
of the energy distribution, that describe universal and non-universal
features of its decaying component.

%%%%%%%%%%%%%%%%%%%%%%%%%%%%%%%%%%%%%%%%%%%%%%%%%%%%%%%%%%%%%%%%
%%%%%%%%%%%%%%%%%%%%%%%%%%%%%%%%%%%%%%%%%%%%%%%%%%%%%%%%%%%%%%%%
\sect{Modeling}
We analyze two models whose dynamics is generated by a RMT Hamiltonian
${\mathcal{H}=\mathcal{H}_0+V}$, with
${\mathcal{H}_0=\mbox{diag}\{E_n\}}$ and $n\in{\mathbb Z}$. The
first one is the Friedrichs model (FM) \cite{friedrichs}, where the
distinguished energy level $E_0$ is coupled to the rest of the
levels $E_{n \neq 0}$ by a rank two matrix. The second one is the
generalized Wigner model (WM) \cite{wigner}, where the perturbation
$V$ does not discriminate between the levels, and is given by a
banded random matrix.
In both cases the system is prepared initially in the eigenstate
corresponding to $E_0$, and the coupling to the other levels is
characterized by the spectral function
\be{-1} 
\nonumber 
\tilde{C}(\omega) 
&=& -{\rm Im}\, 
\Big\langle E_0
\Big|V\left(E_0{+}\omega{-}\tilde{\mathcal{H}}_0{+}i0\right)^{-1}V\Big|E_0\Big\rangle 
\\ \label{eq:1}
&=& \sum_{n\neq0} 
{|V_{n,0}|^2} 2\pi\delta(\omega
-(E_n{-}E_0))
\ee
where $\tilde{\mathcal{H}}_0$ is obtained from $\mathcal{ H}_0$ by
removing the $0^{\tbox{th}}$ row and column. 
An RMT averaging over realizations is implicit in the WM case.

Given a physical system the spectral function $\tilde{C}(\omega)$ 
can be determined numerically (see some examples in \cite{brm,dil}) 
and its various features can be understood analytically by analyzing 
the skeleton which is formed by periodic orbits, bouncing orbits and 
taking into account the Lyapunov instability of the motion. 
In this paper we would like to consider the most dramatic possibility 
of having non-Ohmic spectral function which is conventionally modeled as  
\be{2} 
\tilde{C}(\omega) =
2\pi \epsilon^2 |\omega|^{s-1} \eexp{-|\omega|/\omega_c}
\ee
The cutoff frequency~$\omega_c$ defines the 
bandwidth ${b=\varrho\omega_c}$ of~$V_{nm}$, 
where $\varrho$ is the density of states.
In the FM case~$\pm b$ is the furthest reachable 
state (because $n {\ne} 0$ states are not coupled), 
and therefore the size of the matrix is effectively ${N=b+1}$.

The assumed form Eq.(\ref{e2}) for the spectral function $\tilde{C}(\omega)$ 
constitutes the natural generalization of the standard FM and WM. 
By integrating Eq.(\ref{e2}) over~$\omega$ we see that the perturbation $V$ 
is bounded \hide{for the FM} provided ${s>0}$.
The ${s=1}$ case is what we refer to as the flat continuum (Ohmic case),
for which it is well known that both models leads
to the same exponential decay for the survival probability.
For ${s>2}$ the effect of the continuum can be handled
using $1^{\tbox{st}}$ order perturbation theory.
We focus in the ${0<s<2}$ regime and consider the
${s\neq1}$ case for which a non-linear version of the
Wigner decay problem is encountered.

In the numerical simulations we integrate the
Schr\"odinger equation for $c_n(t)=\langle n|\psi(t)\rangle$ 
starting with the initial condition
$c_n=\delta_{n,0}$ at $t{=}0$. 
We use units such that ${\varrho = \hbar = 1}$, 
and consider a sharp bandwidth $b$.
The integration is done using a self-expanding algorithm \cite{wbr}. 
The spreading profile is described by
the distribution $P_t(n)= \overline{|c_n(t)|^2}$,
where the averaging is over realizations of the Hamiltonian.
The survival probability is  $P(t)=P_t(0)$.
The energy spreading is characterized by the
standard deviation $\Delta E_{\tbox{sprd}}(t)=[\sum_n (E_n{-}E_0)^2 P_t(n)]^{1/2}$,
by the median $E_{50\%}=E_0$, and also by the $E_{25\%}$ and $E_{75\%}$ percentiles.
The width of the core component is defined
as ${\Delta E_{\rm core}(t)=E_{75\%}-E_{25\%}}$.

%%%%%%%%%%%%%%%%%%%%%%%%%%%%%%%%%%%%%%%%%%%%%%%%

%%%%%%%%%%%%%%%%%%%%%%%%%%%%%%%%%
\begin{figure}
\begin{center}
\includegraphics[width=0.9\hsize,clip]{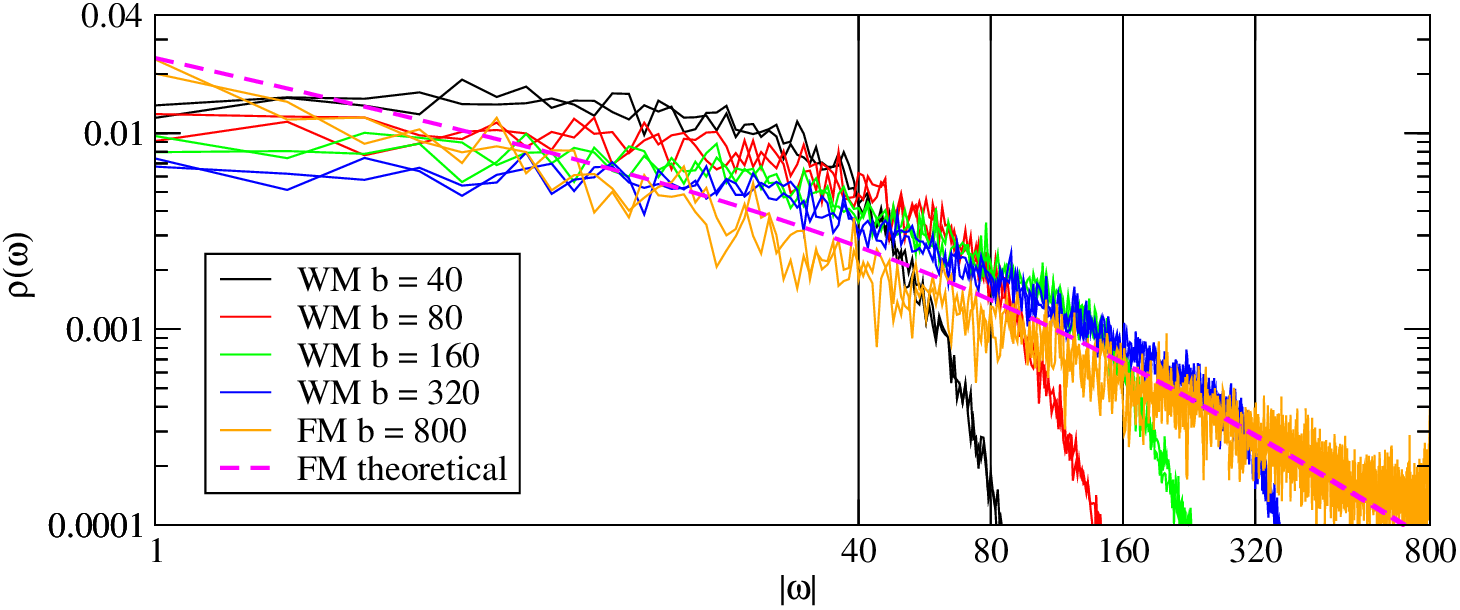}\\
\includegraphics[width=0.9\hsize,clip]{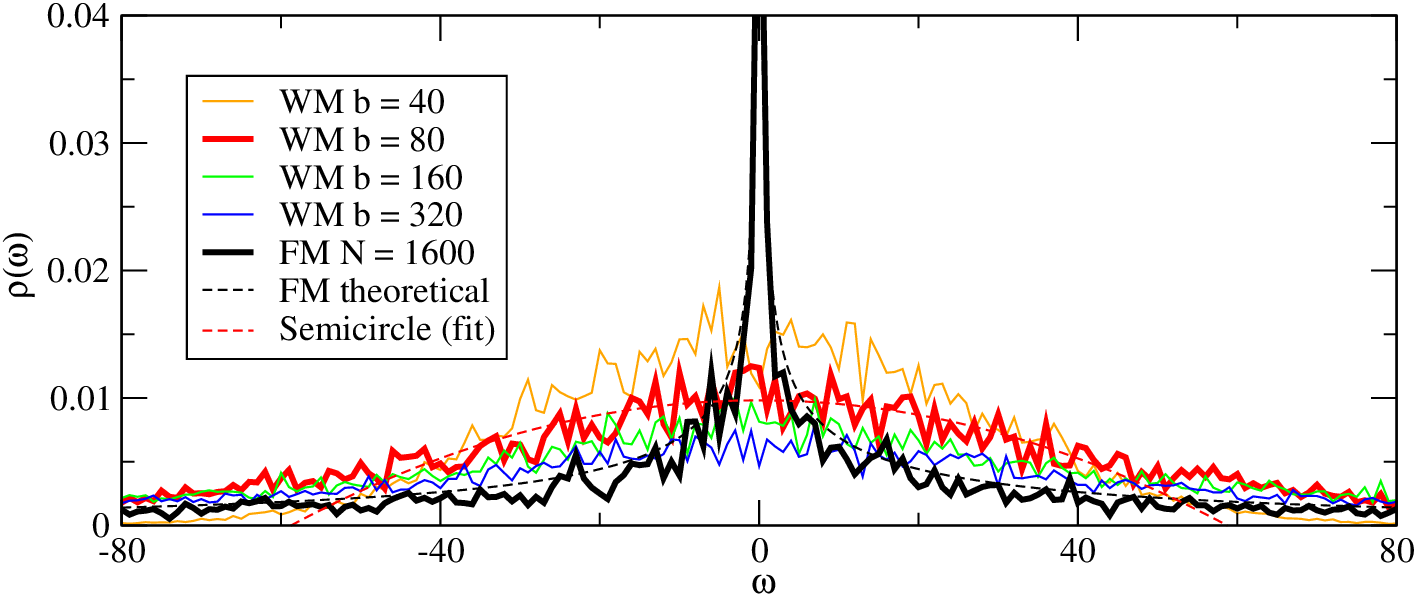}
\end{center}
\caption{LDoS for the FM and for the WM via 
direct diagonalization of $1600\times1600$ matrices 
with ${s=1.5}$ and ${\epsilon=1.44}$.
{\it Upper panel:} The log-log scale 
emphasizes the universality of the tails 
up to $\omega_c$. 
{\it Lower panel:} The log-linear scale emphasizes 
the difference in the non-universal core component.}
\label{fig:ldos}
\end{figure}
%%%%%%%%%%%%%%%%%%%%%%%%%%%%%%%%%%

%%%%%%%%%%%%%%%%%%%%%%%%%%%%%%%%%%%%%%%%%%%%%%%%
%%%%%%%%%%%%%%%%%%%%%%%%%%%%%%%%%%%%%%%%%%%%%%%%
\sect{Time Scales}

A dimensional analysis predicts the existence of 3~relevant time
scales: The Heisenberg time $t_{\tbox{H}}$ which is related to the
density of states $\varrho$; the semiclassical (correlation) time
which is related to the bandwidth~$\omega_c$; and the generalized
Wigner times $t_0$ which is related to the perturbation strength:
\be{0}
t_{\tbox{H}} &=& 2\pi\varrho, 
\ \ \ \ \ \  
t_c = 2\pi/\omega_c
\\
t_0 &=& 
\left(
\frac{2\pi\epsilon^2}
{\bm{\Gamma}(3{-}s) \sin(s\pi/2)}
\right)^{-{1}/{(2-s)}}
\equiv \frac{1}{\gamma_0}
\label{eq5} 
\ee
where $\bm{\Gamma}$ is the Gamma function. 
The numerical prefactor that we have incorporated 
into the definition in Eq.(\ref{eq5}) 
will be explained later in Section.7.
We shall refer to $\varrho^{-1}$  and to $\omega_c$ as the infrared
and ultraviolate cutoffs of the theory. Our main interest is in the
{\em continuum} limit. Assuming further that~$\omega_c$ 
is irrelevant, one expects a decay that is determined by 
the generalized Wigner time~$t_0$. 

It should be clear that the existence of a cutoff free 
universal theory in the continuum limit for ${s\ne1}$
is not self evident. In fact the natural expectation 
might be to have {\em either} infrared {\em or} ultraviolate cutoff 
dependence. Indeed we find that the 2nd moment 
of the spreading depends on the~$\omega_c$ cutoff, 
while~$t_0$ is reflected in the FM case but not in the WM case. 
But as far as $P(t)$ is concerned, we find that 
a one-parameter cutoff free universal theory exists.

%%%%%%%%%%%%%%%%%%%%%%%%%%%%%%%%%%
\begin{figure}
\begin{center}
\includegraphics[width=0.9\hsize,clip]{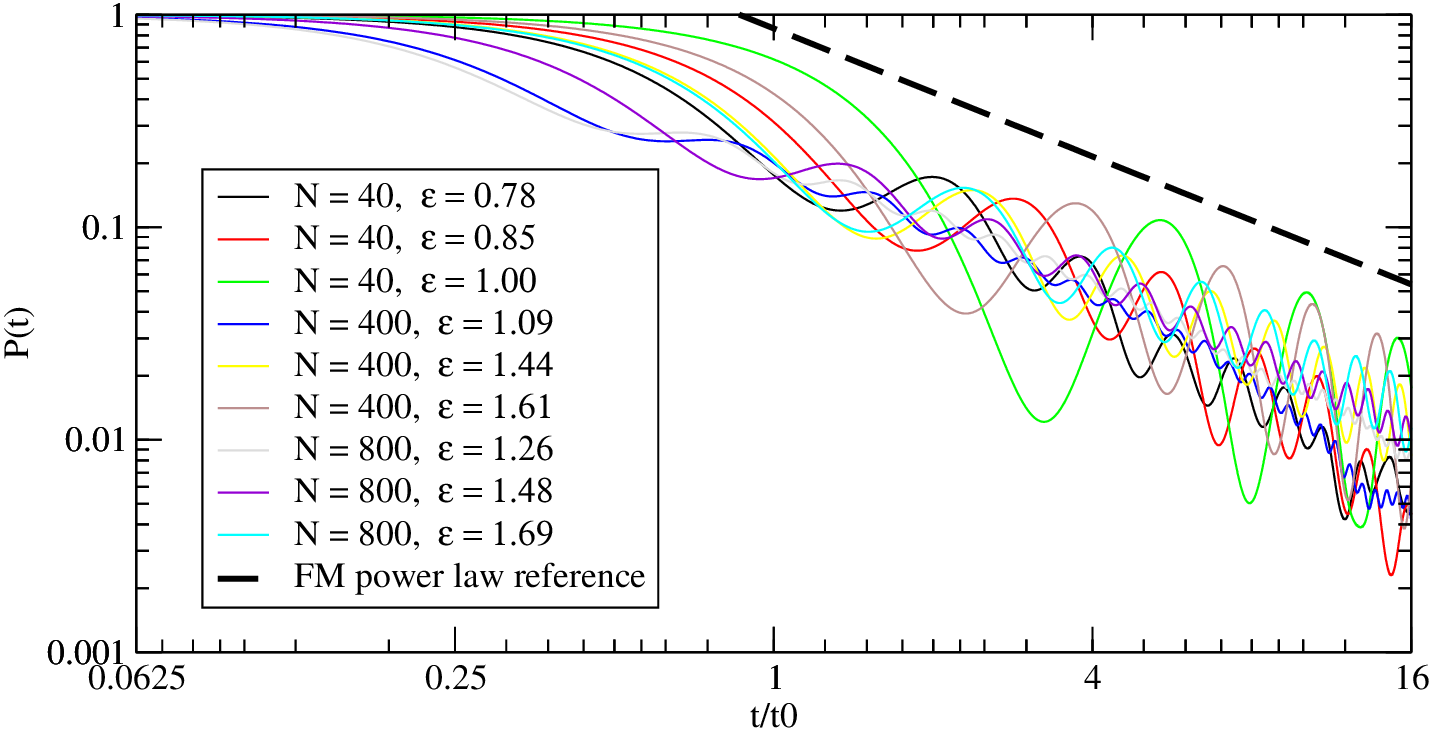}
\includegraphics[width=0.9\hsize,clip]{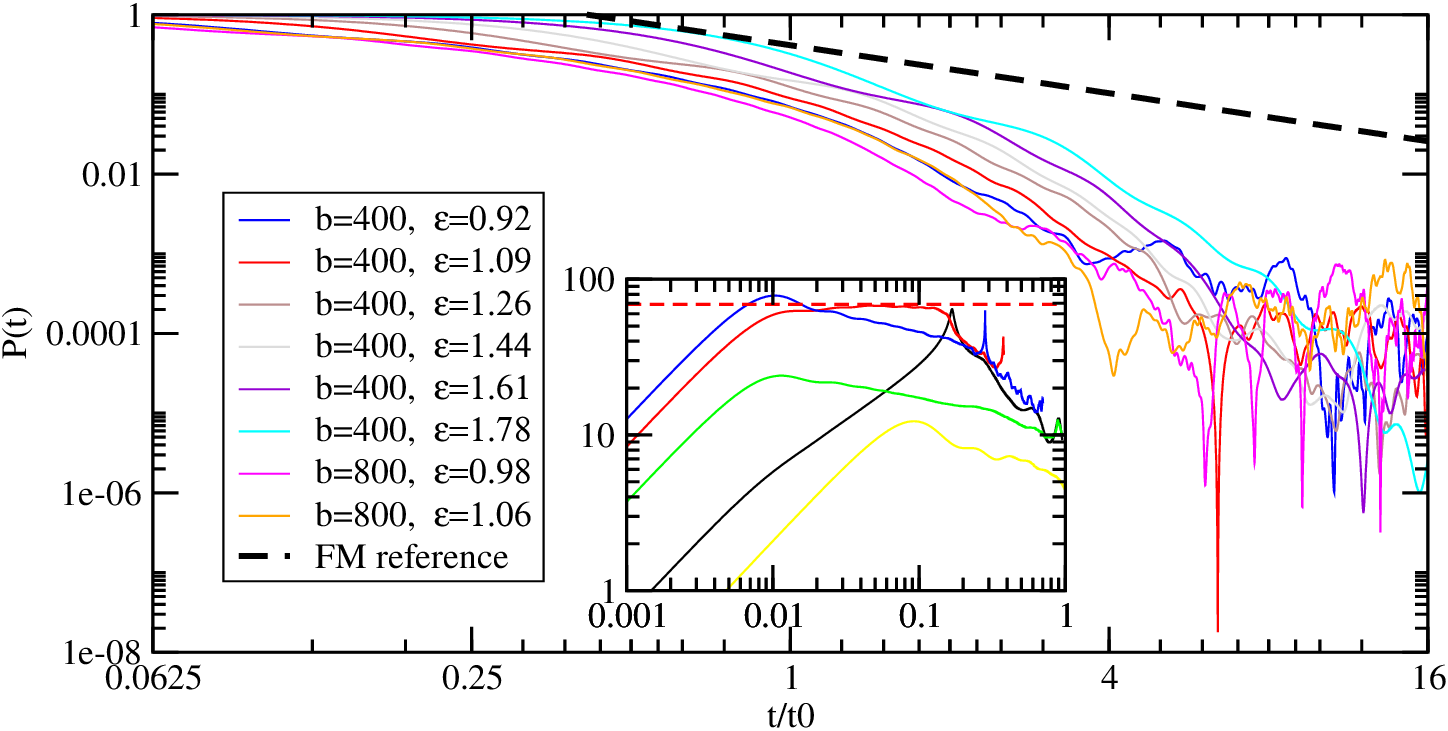}
\end{center}
\caption{The survival probability $P(t)$ for the FM (\textit{top})
and for the WM (\textit{bottom}).
The time is scaled with respect to $t_0$.
For all curves in the main panels ${\varrho=1}$ and ${s=1.5}$.
The WM simulations are presented in log-log scale
in order to contrast it with the FM results.
\textit{Inset:} further analysis displaying $Y=-\ln[P(t)]/t$ vs $X=t$
in a log-log plot for representative runs with $(s,\epsilon)=$ black$(0.30,4.43)$,
red$(1.00,3.24)$, green$(1.25,1.14)$, blue$(1.50,1.09)$, yellow$(1.75,0.50)$, 
showing that the decay in the WM case is described by a stretched exponential.
The red bold dashed line has zero slope, corresponding to simple exponential 
decay for $s{=}1$.} 
\label{fig:survival-prob}
\end{figure}
%%%%%%%%%%%%%%%%%%%%%%%%%%%%%%%%%%

%%%%%%%%%%%%%%%%%%%%%%%%%%%%%%%%%%%%%%%%%%%%%%%%%%%%%%%%%%%
%%%%%%%%%%%%%%%%%%%%%%%%%%%%%%%%%%%%%%%%%%%%%%%%%%%%%%%%%%%
\sect{The LDoS}
Before analyzing the dynamics, it is important to understand the
behavior of the Local Density of States (LDoS) \cite{wigner}, which is defined as follows:
\be{5}
\rho(\omega)=\sum_{\nu} |\langle \nu | 0 \rangle|^2\delta(\omega-(E_{\nu}{-}E_0))
\ee
where $|\nu\rangle$ are the eigenstates of the full Hamiltonian~${\cal H}$. 
An RMT averaging over realizations is implied in the WM case. 
Once the LDoS is computed, we can use it to calculate the survival probability:
\be{0}
\label{eq:7} 
P(t)\ \equiv \ 
\Big|\langle 0|\eexp{-i\mathcal{H}t} | 0 \rangle\Big|^2 
= \left|\mbox{FT}\Big[2\pi \rho(\omega) \Big]\right|^2
\ee
where FT denotes the Fourier transform.
For flat bandprofile (${s=1}$), the LDoS $\rho(\omega)=(1/\gamma_0) f(\omega/\gamma_0)$
is a Lorentzian ${f(x)=(1/\pi)/(1+x^2)}$ \cite{wigner},
leading to a Wigner exponential decay for $P(t)$.
For (${s\ne1}$) the ensuing analysis shows that $\rho(\omega)$ has a
core-tail structure \cite{wls,wbr,brm}. Namely, it consists of two
distinct regions ${x\gg1}$ and ${x<1}$ that reflect universal and
non-universal features of the problem respectively.
The tails ${x\gg1}$ can be calculated using
$1^{\tbox{st}}$ order perturbation theory
leading to ${f(x) \propto 1/x^{3-s}}$.
This component we regard as universal.
The core (${x<1}$) reflects the non-perturbative
mixing of the levels, and it is non-universal.
In the WM case we argue that for ${x \ll 1}$ 
it is semicircle-like, while for FM we 
have a singular behaviour ${f(x) \sim x^{1-s}}$. 
These findings are supported by the numerical
calculations of Fig.1, and are reflected in
the behavior of $P(t)$ as confirmed by the numerical
simulations of Fig.2.

%%%%%%%%%%%%%%%%%%%%%%%%%%%%%%%%%%%%%%%%%%%%%%%%%%%%%
%%%%%%%%%%%%%%%%%%%%%%%%%%%%%%%%%%%%%%%%%%%%%%%%%%%%%
\sect{Friedrichs model}
Using the Schur complement technique, we can calculate analytically
the LDoS for the FM. The Green function is 
$G_{00}(\omega) = \{[\omega{-}\Delta(\omega)]+i (\Gamma(\omega)/2)\}^{-1}$
with the standard notations $\Gamma(\omega)=\tilde{C}(\omega)$, 
\be{0} \nonumber
\Delta(\omega)
&=&
\dashint_{-\infty}^{+\infty}\frac{\widetilde{C}(\omega')}{\omega-\omega'} \ \frac{\mbox{d}\omega'}{2\pi}
\\
&=&
\epsilon^2\pi\cot\left(s{\pi}/{2}\right)|\omega|^{s-1}\mbox{sgn}(\omega)
\ee
In the last line we performed the limit $\omega_c\rightarrow\infty$ 
(with the limiting expression converging in distribution).
The LDoS of Eq.(\ref{e5}) is $-({1}/{\pi}) \im\left[ G_{00}(\omega) \right]$
leading to 
\be{0}
\rho(\omega)
= \frac{1}{\pi}
\,\frac{\Gamma(\omega)/2}{(\omega-\Delta(\omega))^2+(\Gamma(\omega)/2)^2}
\label{eq:rho}
\ee

%%%%%%%%%%%%%%%%%%%%%%%%%%%%%%%%%%%%%%%%%%%%%%%%%%%%%%
%%%%%%%%%%%%%%%%%%%%%%%%%%%%%%%%%%%%%%%%%%%%%%%%%%%%%%
\sect{Wigner Model}
The analysis of the LDOS for the WM can be carried out approximately using a combination
of heuristic and formal methods. Our numerical results reported in Fig.~1 confirm 
that the LDoS has $1^{\tbox{st}}$ order tails ${|V_{n,0}/(E_n-E_0)|^2}$ 
that co-exist with the core (non-perturbative) component. 
We can determine the border $\gamma_0$ between the core and the tail 
simply from the requirement ${p_0 \sim 1}$ where 
\be{0}
p_0 = \int_{\gamma_0}^{\infty} \frac{\tilde{C}(\omega)}{\omega^2} \frac{d\omega}{2\pi} 
\ee
For ${s>2}$ we would have for sufficiently small coupling ${p_0 \ll 1}$
even if we took the limit $\gamma_0 \rightarrow 0$. This means that
$1^{\tbox{st}}$ order perturbation theory is valid as a global approximation.
But for $s<2$ the above equation implies
breakdown of $1^{\tbox{st}}$ order perturbation theory 
at $\gamma_0\sim\epsilon^{2/(2{-}s)}$.
In the tails $\mathcal{H}_0$ dominates over $V$,
while in the core $V$ dominates. Therefore,
as far as the core in concerned, it makes sense to
diagonalize $V$ with an effective cutoff~$\gamma_0$.
Following~\cite{mario}, 
the result for the LDoS lineshape should be semicircle-like,
with width given by the expression
\be{0}
\Delta E_{\tbox{sc}} = \left[\int_0^{\gamma_0} \tilde{C}(\omega) \frac{d\omega}{2\pi} \right]^{1/2}
\ee
where above we use the {\em effective} bandwidth $\gamma_0$,  
which replaces the actual bandwidth $\omega_c$   
(the latter would be appropriate as in~\cite{mario}
if we were considering the WM without the diagonal energies).
The outcome of the integral is ${\Delta E_{\tbox{sc}} \sim \gamma_0}$,  
demonstrating that our procedure is {\em self-consistent}: 
the core has the same width as implied by the breakdown 
of $1^{\tbox{st}}$ order perturbation theory. 
We note that within this perspective the ${s=1}$ Lorentzian is regarded
as composed of a semicircle-like core and $1^{\tbox{st}}$ order tails.

%%%%%%%%%%%%%%%%%%%%%%%%%%%%%%%%%%%%%%%%%%%%%%%%
%%%%%%%%%%%%%%%%%%%%%%%%%%%%%%%%%%%%%%%%%%%%%%%%
\sect{The survival probability}
In the WM case the function $\rho(\omega)$ is {\em smooth} 
with power law tails ${\sim 1/\omega^{1+\alpha}}$ where ${\alpha=2{-}s}$. 
Thanks to the smoothness the FT does not have 
power law tails but is exponential-like. The similarity 
with the $\alpha$-stable Levy distribution suggests 
that $P(t)$ would be similar to a stretched exponential, 
\be{0} 
P(t) \approx& \exp[-(t/t_0)^{2{-}s}] 
\ee
The expression for $t_0$ in Eq.(\ref{eq5}) 
is implied by the observation that $1/|\omega|^{1{+}\alpha}$
tails are FT associated with a discontinuity $-C|t|^{\alpha}$, 
where ${C=[2\bm{\Gamma}(1{+}\alpha)\sin(\alpha\pi/2)]^{-1}}$.

In the FM case we observe that the function $\rho(\omega)$ 
in Eq.(\ref{eq:rho}) features a crossover from
${\omega^{1{-}s}}$ for ${|\omega| \ll\gamma_0}$ 
to ${\Gamma(\omega)/\omega^2}$ for ${|\omega| \gg\gamma_0}$. 
Thus, compared with the WM case,  
the FT has an additional contribution 
from the singularity at $\omega{=}0$, 
and consequently by the Tauberian theorem \cite{sonis}, 
the survival amplitude has a non-exponential 
decay, that for sufficiently long time is described by a power law:  
\be{0}
P(t) \ \ =  \ \ \left|\frac{2\sin((s{-}1)\pi)}{(2{-}s)\pi \ (t/t_0)^{2{-}s}}\right|^2
\ee
The long time behavior is dominated by the non-smooth
feature of the core, and not by the tails. 
Comparing the exponential and the power-law
we can find the expression for the crossover time $t_0'$
that becomes ${t_0'\sim[\log|s{-}1|]^{1/(2{-}s)}t_0 \gg t_0}$
close to the Ohmic limit ($s{\sim}1$).  
For ${s=1}$ only the exponential decay survives. 
We emphasize that the cutoff independent behavior appears 
only after a short transient, i.e. for ${t>t_c}$.   
For completeness we note that for the FM with $s{=}2$
we get ${P(t) \approx |\log(t/t_c')|^2}$, 
that holds for ${t_c<t<t_c'}$ where ${t_c'=t_c\eexp{1/(2\epsilon^2)}}$. 
For ${s{>}2}$ there is an immediate but only partial decay that 
saturates at the value ${P(t)=|1{-}p_0|^2}$ for ${t>t_c}$.

%%%%%%%%%%%%%%%%%%%%%%%%%%%%%%%%%%%%%%%%%%%%%%%
\begin{figure}
\begin{center}
\includegraphics[width=0.9\hsize,clip]{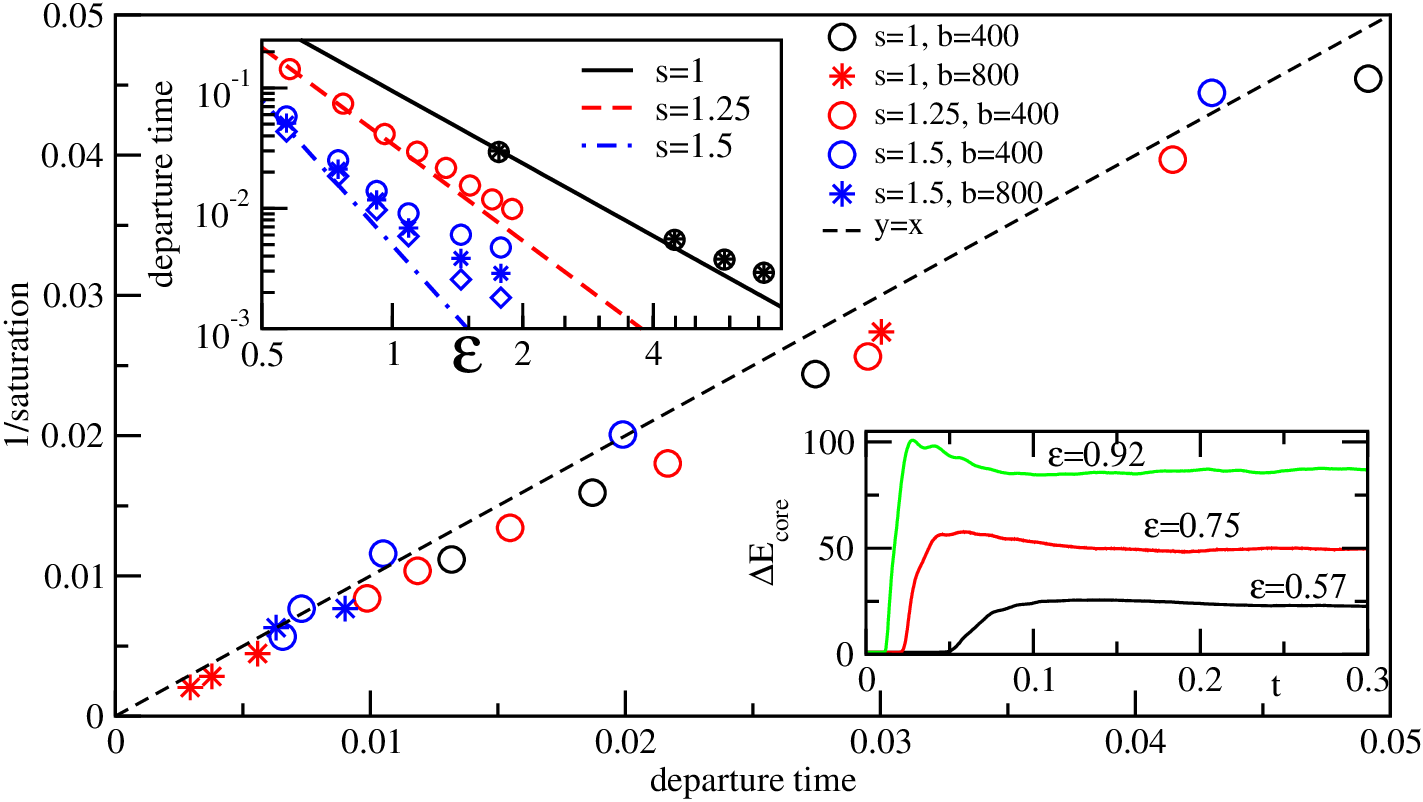}
\end{center}
\caption{
{\em Lower Inset:} Examples for the time evolution of $\Delta E_{\tbox{core}}$ 
for $s{=}1.5$ and $b{=}800$ in the WM case. 
{\em Main panel:} The extracted departure time versus the extracted
inverse saturation value. This scatter diagram demonstrates the validity 
of one parameter scaling.
{\em Upper Inset:} The extracted departure time versus the perturbation strength $\epsilon$.
The theoretical (dashed) lines are based on the $t_0$ estimate of Eq.(\ref{eq5}). The deviations
of the departure time from the theoretical expectation diminish in the limit ${\omega_c \to \infty}$. 
The $\circ$ corresponds to $b=400$, the $\star$ to $b=800$, and the $\diamond$ to $b=1600$.}
\label{fig:n50-analysis}
\end{figure}
%%%%%%%%%%%%%%%%%%%%%%%%%%%%%%%%%%%%%%%%%%%%%%%

%%%%%%%%%%%%%%%%%%%%%%%%%%%%%%%%%
\begin{figure}
\begin{center}
\includegraphics[width=0.8\hsize,clip]{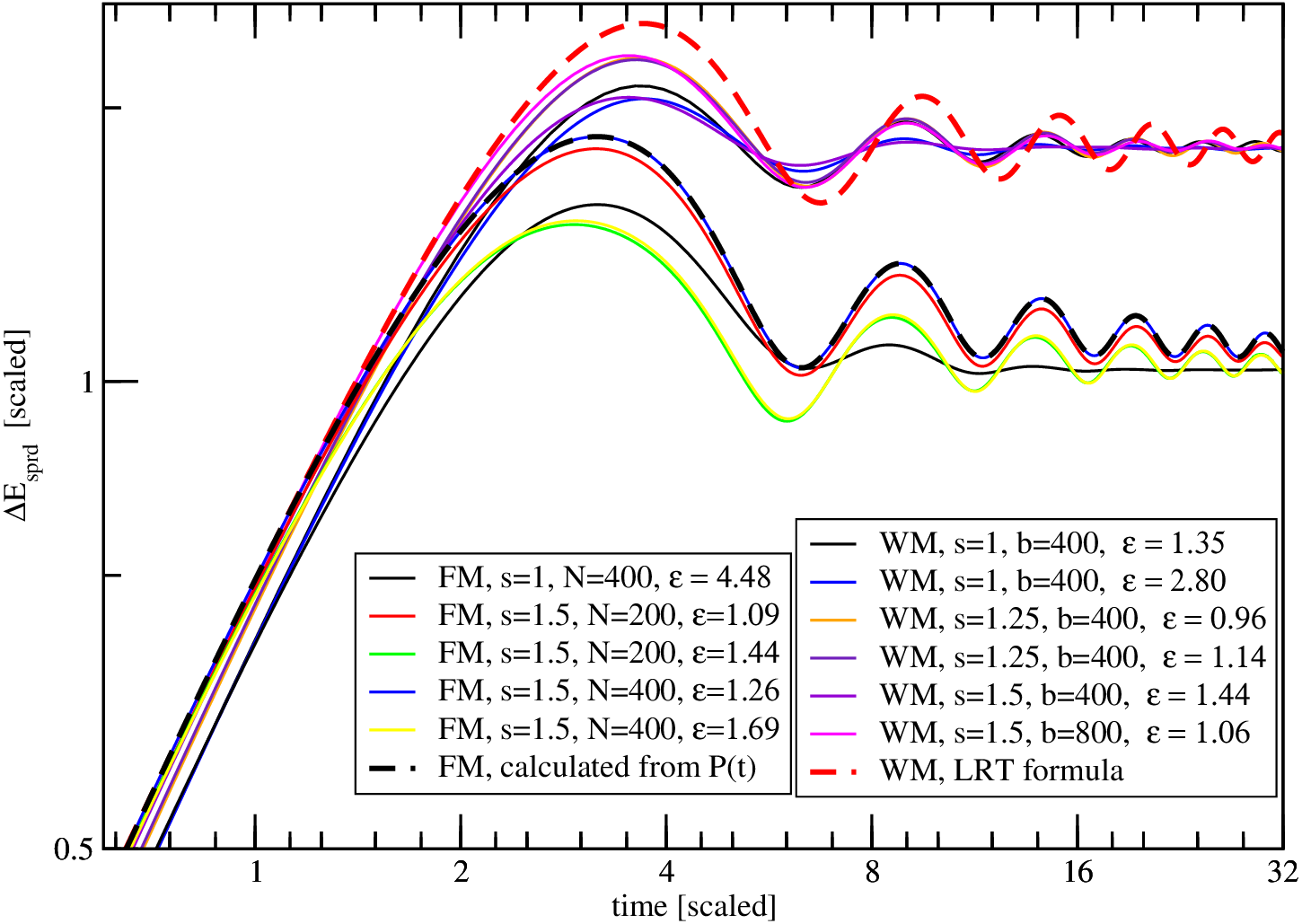}
\end{center}
\caption{
Scaled spread $\Delta E_{\tbox{sprd}}/(\omega_c^s\epsilon^2/s)^{1/2}$
versus scaled time ${\omega_c t}$ for the FM and the WM. 
The linear response theory (LRT) prediction Eq.(\ref{e14}),
for the WM, as well as the exact result Eq.(\ref{e18}) for the FM  
are plotted for comparison.}
\label{fig:tails-spreading}
\end{figure}
%%%%%%%%%%%%%%%%%%%%%%%%%%%%%%%%%%%

%%%%%%%%%%%%%%%%%%%%%%%%%%%%%%%%%%%%%%%%%%%%%%%%%%%%%%%%
%%%%%%%%%%%%%%%%%%%%%%%%%%%%%%%%%%%%%%%%%%%%%%%%%%%%%%%%
\sect{Spreading}
The distinction between core and tail components
becomes physically transparent once we analyze
the time dependent energy spreading of the wavepacket.
Using the same time dependent analysis as in
the ${s=1}$ case of Ref.\cite{wbr}, it is straightforward
to show that the rise of  $\Delta E_{\tbox{core}}(t)$
is at $t\sim t_0$, and its saturation value
is ${\sim \gamma_0}$. Thus $\Delta E_{\tbox{core}}$
should exhibit one parameter scaling with respect to~$t_0$.
In Fig.3 we present the results of the numerical analysis.
Our data, indicate that the expected one-parameter scaling 
is obeyed. We have verified that the slight deviation (shown in the inset) 
from the expected $\epsilon$ dependence is an artifact 
due to having finite (rather then infinite) bandwidth 
in the numerical simulation.

The physics of $\Delta E_{\tbox{sprd}}$ is quite
different and not necessarily universal,
because the second moment is dominated by the tails,
and hence likely to depend on the cutoff $\omega_c$
and diverge in the limit $\omega_c\rightarrow\infty$.
Indeed in the WM case we can use the Linear Response Theory (LRT)
result of \cite{wbr,brm}
\be{14}
\Delta E_{\tbox{sprd}}(t) = \left[ 2\Big(C(0)-C(t)\Big)\right]^{1/2}
\ee
where $C(t)$ is the inverse FT of $\tilde{C}(\omega)$.
This gives the saturated value $(2\omega_c^s\epsilon^2/s)^{1/2}$
as soon as ${t>t_c}$.
We now turn to the FM case. The solution of the 
Schr\"odinger equation for $c_n(t)$ is well known \cite{CDG92}, 
and (setting $E_0{=}0$) can be expressed using the real amplitude $c(t){\equiv}c_0(t)$. 
In particular $P(t)=|c(t)|^2$ and also the energy spreading 
can be computed in a closed form, with the end result 
\be{18} 
\Delta E_{\tbox{sprd}}(t) =
\Big[(1{+}c^2(t))C(0)-\dot{c}(t)^2+2c(t)\ddot{c}(t)\Big]^{1/2}
\ \ \ \ 
\ee
For $t<t_0$ we can use the 
estimates ${c(t)\approx 1}$ and ${\dot{c}(t)\approx 0}$ 
and ${\ddot{c}(t)\approx -C(t)}$ to conclude  
that  $\Delta E_{\tbox{sprd}}(t)$ behaves as in Eq.(\ref{e14}). 
But for ${t>t_0}$ we get %%%% ${\Delta E_{\tbox{sprd}}(t) \approx [(1{+}P(t))C(0)]^{1/2}}$, 
\be{19} 
\Delta E_{\tbox{sprd}}(t) \approx \Big[(1{+}P(t))C(0)\Big]^{1/2}
\ee
leading to a  saturation value smaller 
by factor $\sqrt{2}$, reflecting the non-stationary decay
of the fluctuations as a function of time.
More interestingly Eq.(\ref{e18}) contains a cutoff independent
term that reflects the universal time scale $t_0$.
The numerical results in Fig.4 confirm the validity
of the above expressions. 
We note that in the FM case the effect of recurrences
is more pronounced, because they are better synchronized:
all the out-in-out traffic goes exclusively through the initial state.

%%%%%%%%%%%%%%%%%%%%%%%%%%%%%%%%%%%%%%%%%%%%%%%%%%%%%%%%
%%%%%%%%%%%%%%%%%%%%%%%%%%%%%%%%%%%%%%%%%%%%%%%%%%%%%%%%
\sect{Summary and Discussion}

In this work we have compared two models
that have the same spectral properties,
but still different underlying dynamics.
One of them has an integrable dynamics (FM)
while the other is an RMT type (WM).
This is complementary to our past work \cite{lds}
where we have contrasted a physical model
with its RMT counterpart.

Non-Ohmic coupling to the continuum emerges in various frameworks in physics. 
The general WM analysis might be motivated by the study   
of quantized chaotic systems that exhibit non-Ohmic fluctuations 
due to semi-classically implied long time power-law correlations. 
In fact typical power spectra are in general 
not like ``white noise" (e.g. \cite{wls,lds,brm}).
The general FM analysis might be motivated 
by studies of bound states that are embedded 
in the continuum as in the single-level Fano-Anderson model, 
with diverse realizations in the molecular / atomic / electronic 
context and also with implication regarding photonic lattices:   
see \cite{longhi} and further references therein.
   
It should be clear that by considering 
two special models, we do not cover
the full range of possibilities: In realistic 
circumstances the perturbation might have any rank, 
and there might be non-trivial correlations
between off-diagonal elements 
(which was in fact the case in \cite{lds}). 
Still our results, since they relate to two extreme 
limiting models (FM,WM), serve to illuminate 
the limitations on the universality of Wigner's theory.
 
In the non-Ohmic decay problem that we
have considered a universal generalized Wigner
time scale has emerged. It is not this
time scale but rather the functional form
of the decay that reflects the non-universality.
We find that for ``non-Ohmic chaos" (WM case) 
the survival probability becomes a stretched exponential
beyond the Wigner time scale, 
which is both surprising and interesting.
This is contrasted with the ``integrable"
power-law decay that takes over in the long time limit (FM case), 
and obviously very different from the Ohmic exponential result.
Only the standard case of flat (Ohmic) bandprofile
is fully universal.

It is worth mentioning that in a bosonic second quantized
language the decay of the probability can be re-interpreted  
as the decay of the site occupation~$\hat{n}$.
If the interaction between the bosons is neglected 
this reduction is {\em exact} and merely requires an 
appropriate dictionary.  In the latter context each level 
becomes a bosonic site which is formally like an harmonic oscillator, 
and hence the initially empty continuum is regarded as a zero temperature bath. 
Consequently the decay problem is formally re-interpreted  
as a {\em quantum dissipation} problem 
with an {\em Ohmic} ($s{=}1$) or non-Ohmic ($s{\ne}1$) bath. 
The time scale $t_0$ is associated with the damped motion 
of the generalized coordinate~$\hat{n}$. 
Optionally~$P(t)$ could be related to dephasing, and in this case~$t_0$ 
is reinterpreted as the coherence time, as in Landau's Fermi liquid theory.

%%%%%%%%%%%%%%%%%%%%%%%%%%%%%%%%%%%%%%%%%%%%%%%%%%%%%%%%
%%%%%%%%%%%%%%%%%%%%%%%%%%%%%%%%%%%%%%%%%%%%%%%%%%%%%%%%

\ack
This research is supported by the US-Israel Binational Science Foundation (BSF)

%%%%%%%%%%%%%%%%%%%%%%%%%%%%%%%%%%%%%%%%%%%%%%%%%%%%%%%%
%%%%%%%%%%%%%%%%%%%%%%%%%%%%%%%%%%%%%%%%%%%%%%%%%%%%%%%%%%%

%%%%%%%%%%%%%%%%%%%%%%%%%%%%%%%%%%%%%%%%%%%%%%%%%%%%%%%%%%%%%%%%

\ \\ \ \\ \ \\

{\bf Note after publication:-- } This 2009 arXiv submission 
has been published in J. Phys. A {\bf 43}, 095301 (2010). 
A follow up that contains some more recent additional results, 
and full derivations that were not included in this short report 
is available [arXiv:1003.1645], and has been published 
in Phys. Rev. E {\bf 81}, 036219 (2010). There is another 
follow up regarding ``Quantum anomalies and linear response theory" [arXiv:1003.3303].

\clearpage

%%%%%%%%%%%%%%%%%%%%%%%%%%%%%%%%%%%%%%%%%%%%%%%%%%%%%%%%%%%%%%%%%%%%%%

\Bibliography{99}


\bibitem{AZ02}
N. Auerbach, V. Zelevinsky, Phys. Rev. C {\bf 65}, 034601 (2002);
V.V. Sokolov, V.G. Zelevinsky, Nucl. Phys. A {\bf 504}, 562 (1989).

\bibitem{CDG92}
C. Cohen-Tannoudji, J. Dupont-Roc, G. Grynberg,
{\it Atoms-Photon Interactions: Basic Processes and Applications} (Wiley, New-York, 1992).

\bibitem{NC00}
M.A. Nielsen and I.L. Chuang,
{\emph Quantum computation and quantum information}
(Cambridge University Press,2000).

\bibitem{PAEI94}
V.N. Prigodin, B.L. Altshuler, K.B. Efetov, S. Iida, Phys. Rev. Lett. {\bf 72}, 546 (1994);
B.L. Altshuler et al., Phys. Rev. Lett. {\bf 78}, 2803 (1997).


\bibitem{BH91}
C. W. J. Beenakker, H. van Houton,
in {\it Solid State Physics: Advances in Research and Applications},
Ed. H. Ehrenreich and D. Turnbull, 1-228 {\bf 44} (Academic Press, New York, 1991).


\bibitem{PRSB00}
E. Persson, I. Rotter, H.-J. St\"ockmann,
M. Barth, Phys. Rev. Lett. {\bf 85}, 2478 (2000).

\bibitem{WW30}
V. Weisskopf and E.P. Wigner,
Z. Phys. {\bf 63}, 54 (1930).

\bibitem{izrailev}
F.M. Izrailev, A.Castaneda-Mendoza, Phys. Lett. A {\bf 350}, 355 (2006);
V.V. Flambaum, F.M.Izrailev, Phys. Rev. E {\bf 64} 026124 (2001);
V.V. Flambaum, F.M.Izrailev, Phys. Rev.E {\bf 61}, 2539 (2000).

\bibitem{fyodo}
Y.V. Fyodorov, O.A. Chubykalo, F.M. Izrailev, G. Casati, Phys. Rev. Lett. {\bf 76}, 1603 (1996).

\bibitem{GACMP97}
J.L. Gruver et al.,
Phys. Rev E {\bf 55}, 6370 (1997).

\bibitem{S01}
P.G. Silvestrov, Phys. Rev. B {\bf 64}, 113309 (2001); 
A. Amir, Y. Oreg, Y. Imry, Phys. Rev. A {\bf 77}, 050101(R) (2008).

\bibitem{brm} 
M. Hiller, D. Cohen, T. Geisel and T. Kottos,
Annals of Physics {\bf 321}, 1025 (2006).

\bibitem{friedrichs}
K.O. Friedrichs,
Comm. Pure  Appl.  Math. \ {\bf 1}, 361 (1948).

\bibitem{wigner}
E. Wigner,
Ann. Math {\bf 62} 548 (1955);
{\bf 65} 203 (1957).

\bibitem{dil}
A. Barnett, D. Cohen and E.J. Heller, 
Phys. Rev. Lett. {\bf 85}, 1412 (2000); \ J. Phys. A {\bf 34}, 413-437 (2001).

\bibitem{wls} 
D. Cohen, E.J. Heller, 
Phys. Rev. Lett. {\bf 84}, 2841 (2000).  

\bibitem{wbr} 
D. Cohen, F.M. Izrailev, T. Kottos, 
Phys. Rev. Lett. {\bf 84}, 2052 (2000).
T. Kottos and D. Cohen, Europhys. Lett. {\bf 61}, 431 (2003).

\bibitem{mario} 
M. Feingold, Europhysics Letters {\bf 17}, 97 (1992).

\bibitem{sonis}
K.~Soni, R.P.~Soni, J. Math. Anal. Appl. {\bf 49}, 477 (1975).

\bibitem{lds} 
D. Cohen and T. Kottos, 
Phys. Rev. E {\bf 63}, 36203 (2001).

\bibitem{longhi} 
S. Longhi, Phys. Rev. Lett. {\bf 97}, 110402 (2006);
Eur. Phys. J. B {\bf 57}, 45 (2007).

\end{thebibliography}
\end{document}